\documentclass[aps,prb,twocolumn,showpacs,superscriptaddress,longbibliography]{revtex4-2}
\usepackage{amsfonts}
\usepackage{amsmath}
\usepackage{graphicx}
\usepackage{bm}
\usepackage{soul}
\usepackage{amssymb}
\usepackage{dcolumn}
\usepackage{color}
\usepackage{multirow}
\usepackage{booktabs}
\usepackage{latexsym}
\usepackage[colorlinks,
linkcolor=blue,
anchorcolor=blue,
citecolor=blue,
urlcolor=blue,
]{hyperref}

\begin{document}

    \title{Enhancement of temperature of quantum anomalous Hall effect in two-dimensional germanene/magnetic semiconductor 
    heterostructures}

    \author{Qing-Han Yang}
    \affiliation{Kavli Institute for Theoretical Sciences, University of Chinese Academy of Sciences, Beijing 100049, China}

    \author{Jia-Wen Li}
    \affiliation{Kavli Institute for Theoretical Sciences, University of Chinese Academy of Sciences, Beijing 100049, China}

    \author{Xin-Wei Yi}
    \affiliation{School of Physical Sciences, University of Chinese Academy of Sciences, Beijing 100049, China}

    \author{Xiang Li}
    \affiliation{Kavli Institute for Theoretical Sciences, University of Chinese Academy of Sciences, Beijing 100049, China}

    \author{Jing-Yang You}
    \affiliation{Peng Huanwu Collaborative Center for Research and Education, Beihang University, Beijing 100191, China}

    \author{Gang Su}
    \email{gsu@ucas.ac.cn}
    \affiliation{Kavli Institute for Theoretical Sciences, University of Chinese Academy of Sciences, Beijing 100049, China}
    \affiliation{School of Physical Sciences, University of Chinese Academy of Sciences, Beijing 100049, China}
    \affiliation{Institute of Theoretical Physics, Chinese Academy of Sciences, Beijing 100190, China}
    \affiliation{Physical Science Laboratory, Huairou National Comprehensive Science Center, Beijing 101400, China}

    \author{Bo Gu}
    \email{gubo@ucas.ac.cn}
    \affiliation{Kavli Institute for Theoretical Sciences, University of Chinese Academy of Sciences, Beijing 100049, China}
    \affiliation{Physical Science Laboratory, Huairou National Comprehensive Science Center, Beijing 101400, China}

    \begin{abstract}
        Quantum anomalous Hall effect (QAHE) is significant for future low-power electronics devices, where a main challenge 
        is realizing QAHE at high temperatures. In this work, based on experimentally reported two-dimensional (2D) germanene 
        and magnetic semiconductors Cr$_2$Ge$_2$Te$_6$ and Cr$_2$Si$_2$Te$_6$, and the first principle calculations, 
        germanene/magnetic semiconductor heterostructures are investigated. Topologically nontrivial edge states 
        and quantized anomalous Hall conductance are demonstrated. It is shown that the QAHE temperature can be enhanced 
        to approximately 62 K in germanene/monolayer (ML) Cr$_2$Ge$_2$Te$_6$ with 2.1\% tensile strain, 64 K in 
        germanene/bilayer (BL) Cr$_2$Ge$_2$Te$_6$ with 1.4\% tensile strain, and 50 K in germanene/ML Cr$_2$Si$_2$Te$_6$ 
        with 1.3\% tensile strain. With increasing tensile strain of these heterostructures, the band gap decreases 
        and the Curie temperature rises, and the highest temperature of QAHE is obtained. Since these 2D materials were 
        discovered in recent experiments, our results provide promising materials for achieving high-temperature QAHE. 
    \end{abstract}
    \pacs{}
    \maketitle

    %%%%%%% Main text %%%%%%%%%%%%%%%%%%%%%

    %%%%%%%%% Section1
    \section{Introduction}
    Quantum anomalous Hall effect (QAHE) is one of the most exciting phenomenon of matter and is characterized by an integer Chern 
    number $\mathcal{C}$ \cite{Haldane1988a,thouless1982}. A key manifestation of QAHE is the dissipationless chiral edge current. 
    It has the potential to revolutionize future low-power quantum electronic devices. 
    Therefore, much theoretical and experimental research has been conducted to achieve QAHE at high temperatures 
    \cite{chang2013,chang2015,checkelsky2014,feng2015,grauer2015,kandala2015,kou2014,kou2015,mogi2015,deng2020,liu2020,ge2020,hu2020,tian2020,you2021,you2022,you2023,watanabe2019,serlin2020,sharpe2019,tschirhart2021,liQ2021,qiu2023,zhu2020,serlin2020,sharpe2019,tschirhart2021,liQ2021}. 
    Various methods are being explored to achieve QAHE. Introducing ferromagnetic order in topological insulators (TIs) is a primary 
    method for achieving QAHE. The first observation of QAHE occurred in magnetically doped TIs, i.e., 
    Cr-doped (Bi,Sb)$_2$Te$_3$ thin films at 30 mK \cite{chang2013}. Subsequent research has focused on increasing the QAHE 
    temperature. For example, the anomalous Hall resistance $\rho_{xy}$ reaches a maximum value of $0.95 h/e^2$ in a 10 quintuple-layer 
    (QL) thin film of Cr$_{0.1}$(Bi$_{0.5}$,Sb$_{0.5}$)$_{1.9}$Te$_3$ at 280 mK \cite{kandala2015}, and a six QL 
    (Cr$_{0.12}$Bi$_{0.26}$Sb$_{0.62}$)$_2$Te$_3$ film exhibits a quantized anomalous Hall resistance $R_{yx}$ up to 0.3 K \cite{kou2015}. 
    To date, the highest reported record QAHE temperature in these doped magnetic TIs is approximately 2 K \cite{mogi2015}. 
    Because disordered magnetic impurities can inevitably degrade sample quality, thus limiting the QAHE temperature. 
    Many researchers are investigating intrinsic QAHE materials. In recent years, experimental breakthroughs in QAHE have been 
    achieved in the van der Waals (vdW) layered material MnBi$_2$Te$_4$ \cite{deng2020,liu2020,ge2020}. MnBi$_2$Te$_4$ is a 
    layered magnetic TI consisting of Te-Bi-Te-Mn-Te-Bi-Te septuple layers (SLs) with antiferromagnetic ground state characterized 
    by antiferromagnetic interlayer coupling and ferromagnetic intralayer coupling. In five SL flakes, the quantization temperature 
    can be increased to 6.5 K by aligning all layers ferromagnetically with an external magnetic field \cite{deng2020}. 
    Recent studies have revealed the QAHE in moir\'e materials can be derived from both graphene 
    \cite{serlin2020,sharpe2019,tschirhart2021} and transition-metal dichalcogenides (TMDs) \cite{liQ2021}. In these systems, 
    QAHE originates from spontaneous valley polarization. However, the typically small band gap of such systems imposes significant 
    limitations on the realization temperature of the QAHE \cite{qiu2023,zhu2020}. 

    \par Another approach to achieving the QAHE state involves fabricating TI/magnetic semiconductor heterostructures. 
    Recent experimental evidnece has shown that low-buckled epitaxial germanene is a quantum spin Hall insulator with 
    a large bulk gap and robust metallic edges at 77 K \cite{germanene2023}. The quantized spin Hall conductance has also 
    been confirmed by first principle calculations \cite{matusalem2019}. Furthermore, two-dimensional (2D) 
    Cr$_2$Ge$_2$Te$_6$ has been experimentally demonstrated as a ferromagnetic semiconductor, with a Curie temperature 
    ($T_C$) of 28 K for bilayer (BL) Cr$_2$Ge$_2$Te$_6$ \cite{gong2017}. Previous theoretical studies suggest that chiral edge 
    states and quantized Hall conductance can be realized in germanene/Cr$_2$Ge$_2$Te$_6$ heterostructures \cite{zou2020,zhang2019}. 
    A key question is whether the QAHE temperature can be enhanced in germanene/monolayer (ML) Cr$_2$Ge$_2$Te$_6$. Recent 
    studies have indicated that tensile strain is an effective method for increasing the $T_C$ of 2D Cr$_2$Ge$_2$Te$_6$ 
    \cite{dong2020,oneill2023}. First-principles calculations predict a $T_C$ of 144 K for ML Cr$_2$Ge$_2$Te$_6$ under 5\% 
    tensile strain \cite{oneill2023}. Experimentally, strain can be applied to samples through various techniques 
    \cite{ni2021,krizman2024,mohiuddin2009,oneill2023,siskins2022,zeng}, including using a micro-manipulator on the substrate 
    to transfer strain to the sample \cite{ni2021} or exploiting lattice mismatch between different materials \cite{krizman2024}. 
    Gradient-modulated strain up to 4\% has been reported in experiments with vdW heterostructures \cite{zeng}. 

    \par In this paper, we report our first-principles study on enhancing the QAHE temperature in germanene/magnetic 
    semiconductor heterostructures. We investigated the effect of biaxial tensile strain on these heterostructures to increase 
    the QAHE temperature. Our calculations show that a QAHE temperature of 62 K is achieved in the germanene/ML 
    Cr$_2$Ge$_2$Te$_6$ heterostructure with 2.1\% tensile strain. For the germanene/BL Cr$_2$Ge$_2$Te$_6$ heterostructure, 
    a QAHE temperature of 64 K is obtained with 1.4\% tensile strain, while the germanene/ML Cr$_2$Si$_2$Te$_6$ heterostructure 
    exhibits a QAHE temperature of 50 K under 1.3\% tensile strain. These results suggest a promising route for achieving 
    high-temperature QAHE using experimentally available 2D materials. 

    \begin{figure}[tphb]
    	\centering
        \includegraphics[scale=0.46]{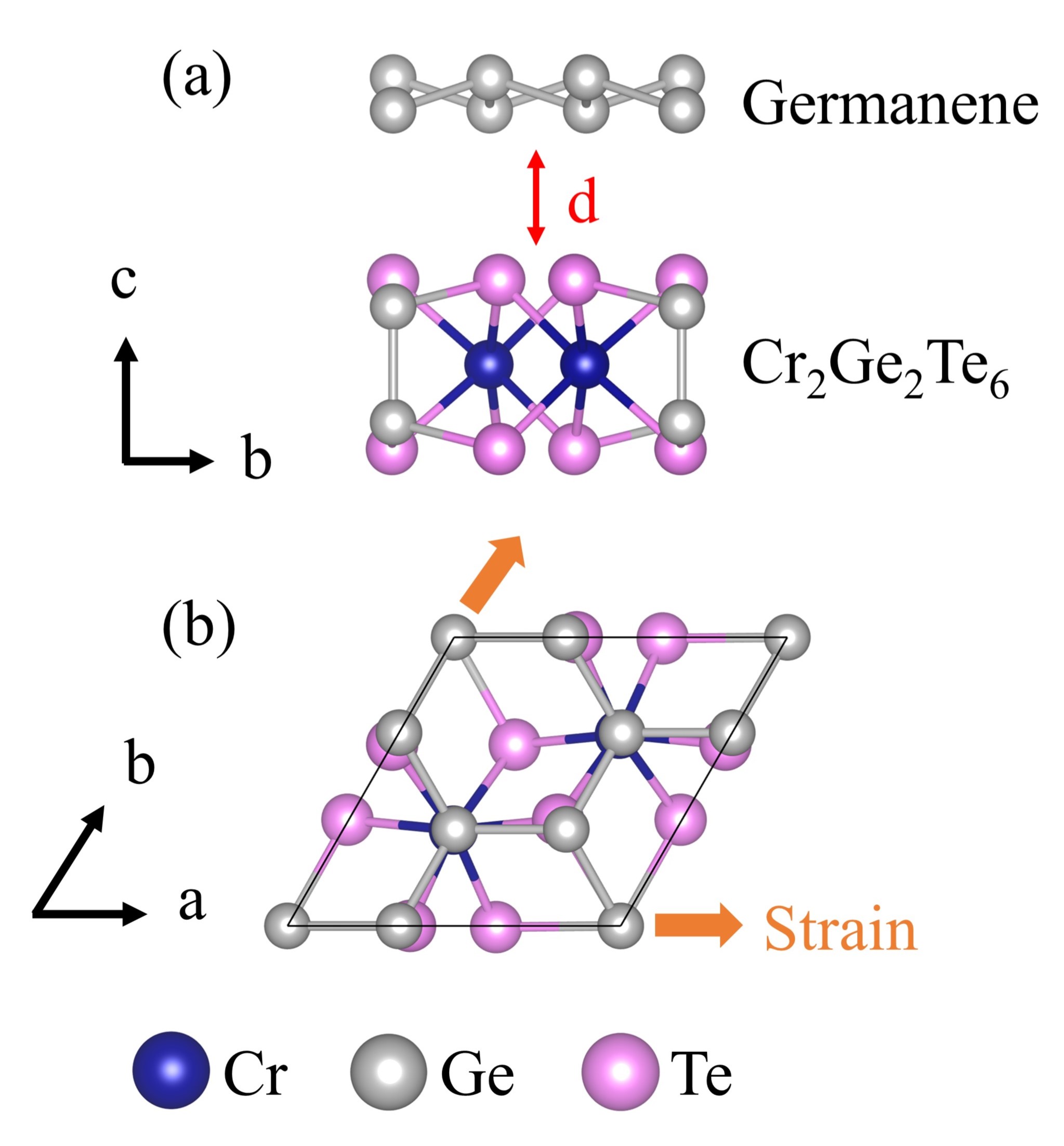}
        \caption{
    	Crystal structures of two-dimensional germanene/ML Cr$_2$Ge$_2$Te$_6$ heterostructure: (a) side view, and 
        (b) top view, respectively. The arrows indicate biaxial tensile strain applied to the heterostructure.
    	}\label{Fig.1}
    \end{figure}
    
    %%%%%%%%% Section2
    \section{Method}
    Our first-principle calculations are based on Vienna $ab$ $initio$ simulation package (VASP) \cite{kresse1996}. We chose the 
    Perdew-Burke-Ernzerhof (PBE) form with the generalized gradient approximation (GGA) to describe the exchange-correlation potential 
    \cite{perdew1996}. The DFT-D3 method is adopted to describe the vdW interaction between germanene and magnetic 
    semiconductor layers \cite{grimme2010}. The vacuum length is taken as 20 \AA, which is enough to isolate the 
    present two-dimensional (2D) system. Spin-orbit coupling (SOC) is taken into account in the calculations. It is reasonable 
    to adopt Hubbard $U = 4$ eV for 3$d$ electrons of Cr in ML Cr$_2$Ge$_2$Te$_6$ \cite{dong2019}. The 
    $12 \times 12 \times 1$ $\Gamma$-center k point is used for the Brillouin zone (BZ) sampling. All the structures were fully 
    relaxed with the convergence precision of energy and force of $10^{-7}$ eV and $10^{-3}$ eV/\AA, respectively. 
    The Wannier90 code was used to construct a tight-binding Hamiltonian \cite{Wannier902008,wannier902014} and WannierTools 
    code was used to obtain topological properties of band \cite{wu2018}.

    %%%%%%%%% Section3
    \section{QAHE in germanene/ML C\lowercase{R}$_2$G\lowercase{E}$_2$T\lowercase{E}$_6$ heterostructure}
    \par The crystal structure with the side and top views of the germanene/ML Cr$_2$Ge$_2$Te$_6$ heterostructure is shown 
    in Fig. \ref{Fig.1}. The heterostructure is constructed using a $\sqrt{3} \times \sqrt{3}$ supercell of germanene 
    and a $1 \times 1$ unit cell of ML Cr$_2$Ge$_2$Te$_6$ with an interlayer distance of $d = 3.45$ \AA. The calculated 
    lattice mismatch between germanene and Cr$_2$Ge$_2$Te$_6$ is less than 1\%. The optimized in-plane lattice constant 
    of the germanene/Cr$_2$Ge$_2$Te$_6$ heterostructure is $a = 6.954$ {\AA}, which is slightly smaller than that of 
    the optimized ML Cr$_2$Ge$_2$Te$_6$ ($a = 6.964$ \AA). 
    
    \begin{figure}[tphb]
    	\centering
        \includegraphics[scale=0.3]{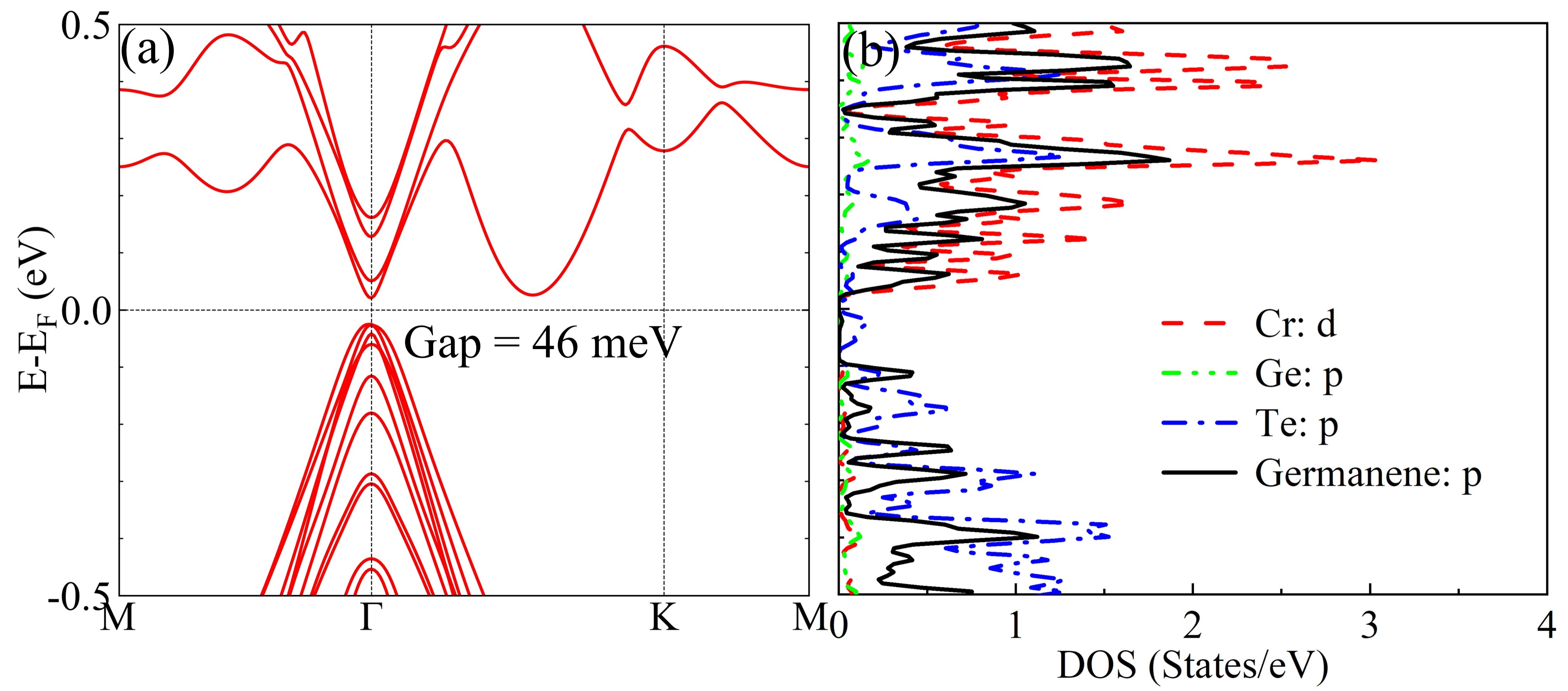}
        \caption{
    	(a) Electronic band structures and (b) density of states of germanene/ML Cr$_2$Ge$_2$Te$_6$ heterostructure 
        obtained by the first principle calculations with SOC. 
    	}\label{Fig.2}
    \end{figure}

    \begin{figure}[tphb]
    	\centering
        \includegraphics[scale=0.46]{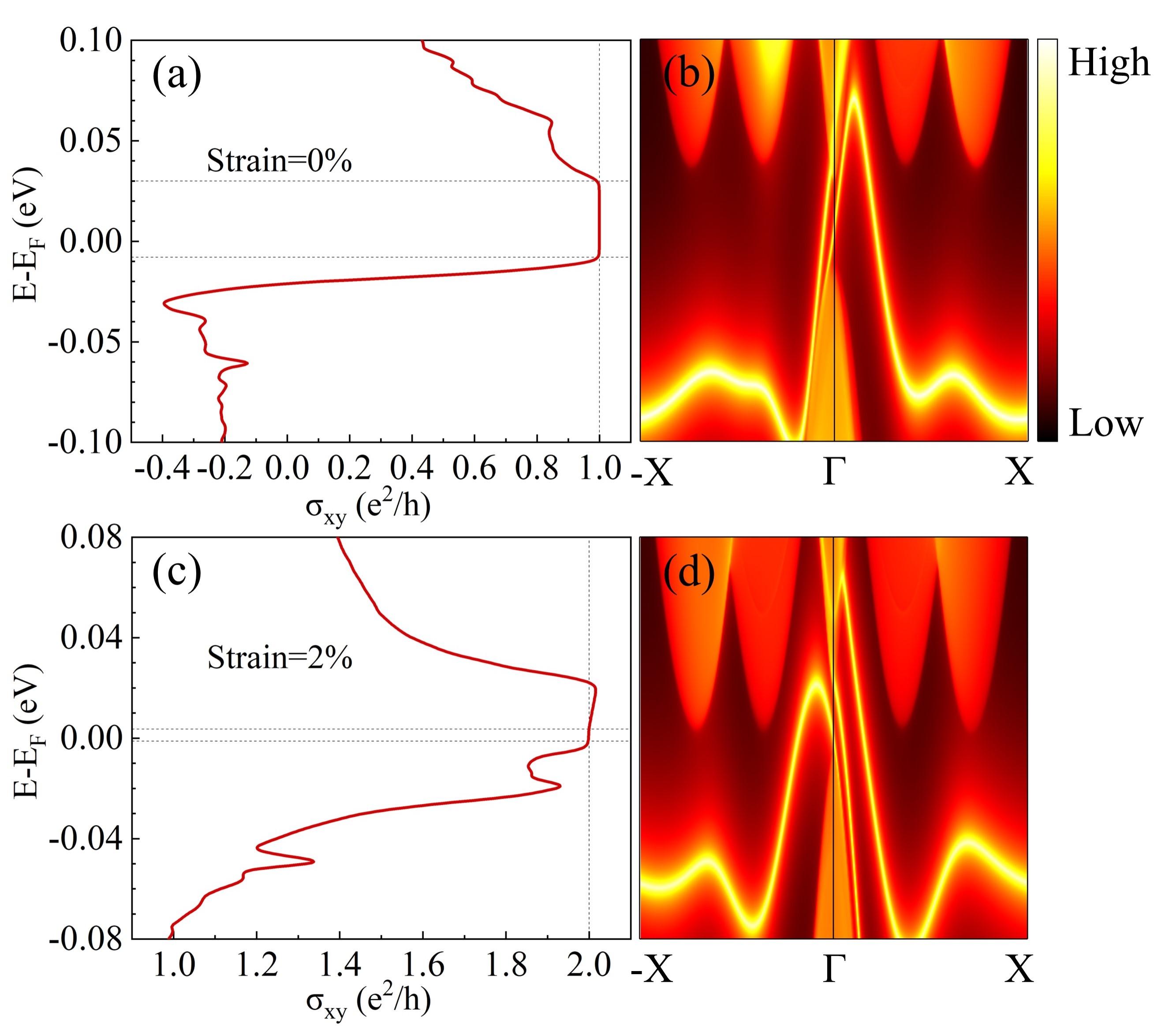}
        \caption{
            For germanene/ML Cr$_2$Ge$_2$Te$_6$ heterostructure, (a) anomalous Hall conductivity and (b) chiral edge states 
            near the Fermi level for the case without strain. (c)-(d) for the case with 2\% tensile strain.  
    	}\label{Fig.3}
    \end{figure}

    \begin{figure*}[tphb!]
    	\centering
        \includegraphics[scale=0.52]{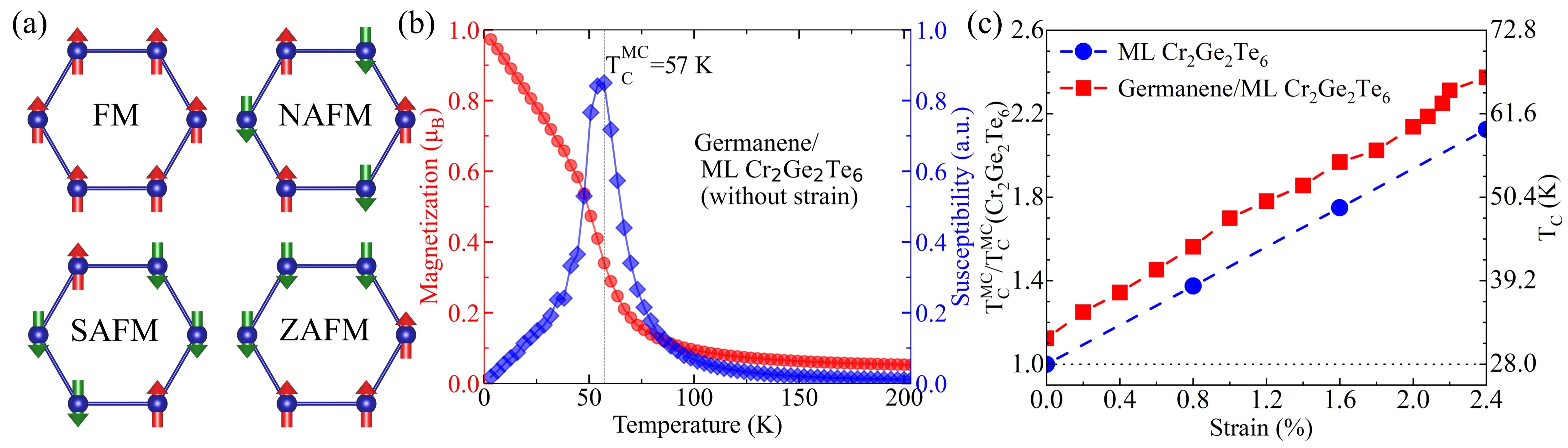}
        \caption{
            (a) Four different spin configurations for Cr atoms in the honeycomb lattice: 
            FM (ferromagnetic), Neel AFM (antiferromagnetic), Stripy AFM and Zigzag AFM. 
            Red and green arrows represent the Cr atom with spin up and spin down polarizations, respectively. 
            (b) The calculated magnetization and susceptibility as a function of temperature for of germanene/Cr$_2$Ge$_2$Te$_6$ 
            heterostructure without applying strain by MC simulations, where the Curie temperature is obtained as $T^{MC}_C$ = 57 K. 
            (c) The rescaled Curie temperature $T_C$ as a function of tensile strain by MC simulations for germanene/ML 
            Cr$_2$Ge$_2$Te$_6$ heterostructure and ML Cr$_2$Ge$_2$Te$_6$. See text for details. 
    	}\label{Fig.4}
    \end{figure*}

    \par The electronic band structures and density of states (DOS) of germanene/ML Cr$_2$Ge$_2$Te$_6$ heterostructure are shown in 
    Fig. \ref{Fig.2}. With SOC included, the germanene/ML Cr$_2$Ge$_2$Te$_6$ heterostructure exhibits a finite band gap of 46 meV. 
    This indicates that even at room temperature, thermal fluctuations are insufficient to fully populate the band gap. 
    The Dirac cone around the $\Gamma$ point is predominantly contributed by $p_z$ orbits of germanene. Notably, this band gap 
    is larger than that of pristine germanene. Since SOC depends on the atomic number $(\sim Z^4)$ and SOC strength of Te is 
    larger than that of germanene, the observed band gap enhancement is attributed to interlayer coupling and hybridization. 
    To investigate the topological properties, maximally localized Wannier functions (MLWFs) implemented in the Wannier90 code 
    were used to fit our density functional theory (DFT) band calculations. The anomalous Hall conductivity 
    is calculated using $\sigma_{xy} = \mathcal{C} \frac{e^2}{h}$, where $\mathcal{C}$ is the Chern number defined as 
    \begin{equation}
        \mathcal{C} = \frac{1}{2\pi} \int_{BZ} {\bf{\Omega}}({\bf{k}}) d {\bf{k}}, 
    \end{equation}
    \noindent \textcolor{red}{and ${\bf{\Omega}}(k)$ is the Berry curvature that can be obtained as } 
    \begin{equation}
        {\bf{\Omega}}({\bf{k}}) = -\sum_{n < E_F} \sum_{m \neq n} 2 Im 
        \frac{<\psi_{n{\bf{k}}}|v_x|\psi_{m{\bf{k}}}><\psi_{m{\bf{k}}}|v_y|\psi_{n{\bf{k}}}>}{(\epsilon_{m{\bf{k}}} - \epsilon_{n{\bf{k}}})^2}, 
    \end{equation}
    \noindent where $\psi_{m(n){\bf{k}}}$ are the Bloch wave functions. $\epsilon_{m(n){\bf{k}}}$ are 
    the eigenvalues and $v_{x(y)}$ are the velocity operators. Such calculation has been implemented in the WannierTools 
    code \cite{wu2018}. 
    The calculated Chern number for the valence band near the Fermi level is $\mathcal{C} = 1$, confirming the topologically nontrivial 
    nature of the band structure. Fig. \ref{Fig.3}(a) shows the calculated anomalous Hall conductance as a function of 
    chemical potential, revealing a quantized Hall plateau at $\sigma_{xy} = \frac{e^2}{h}$. The plateau is consistent 
    with the presence of one topologically nontrivial edge state connecting the bulk states in Fig. \ref{Fig.3}(b). 
    
    \par To investigate magnetic properties of this heterostructure, we considered the following Heisenberg-type Hamiltonian:
    
    \begin{equation}
        \begin{split}
            H = &\sum_{<i,j>} J_1 \textbf{S}_i \cdot \textbf{S}_j + \sum_{<<i,j>>} J_2 \textbf{S}_i \cdot \textbf{S}_j \\
            &+ \sum_{<<<i,j>>>} J_3 \textbf{S}_i \cdot \textbf{S}_j + M \sum_{i} (S^z_i)^2, 
            \label{equ.1}
        \end{split}
    \end{equation}

    \noindent where $J_1$, $J_2$, $J_3$ denote the nearest neighbor, second-nearest-neighbor and third-nearest-neighbor exchange couplings, 
    respectively. To determine the magnetization direction, we preformed total-energy calculations for both  
    out-of plane and in-plane magnetization. Our results show that the easy axis is perpendicular to the 2D plane, 
    consistent with experimental results for 2D Cr$_2$Ge$_2$Te$_6$ \cite{gong2017}. To obtain exchange couplings, we 
    considered four spin configurations: ferromagnetic (FM), Neel antiferromagnetic (AFM), Stripy AFM and Zigzag AFM as shown 
    in Fig. \ref{Fig.4}(a). The energy of four configurations are expressed as 

    \begin{equation}
        \begin{split}
            E_{FM} &= 12 J_1 + 24 J_2 + 12 J_3 + E_0, \\
            E_{NAFM} &= -12 J_1 + 24 J_2 - 12 J_3 + E_0, \\
            E_{SAFM} &= -4 J_1 - 8 J_2 + 12 J_3 + E_0, \\
            E_{ZAFM} &= 4 J_1 - 8 J_2 - 12 J_3 + E_0, 
            \label{equ.2}
        \end{split}
    \end{equation}

    \noindent where $E_0$ is the energy which is independent of the spin configurations. $J_1$, $J_2$, $J_3$ are provided in supplementary 
    materials (SMs) \cite{SMs}. The Monte Carlo (MC) simulations were performed on a 20 $\times$ 20 $\times$ 1 hexagonal lattice with 
    periodic boundary conditions, using $10^6$ MC steps per temperature. The resulting temperature dependence of the magnetization and 
    susceptibility for unstrained germanene/ML Cr$_2$Ge$_2$Te$_6$ heterostructure is presented in Fig. \ref{Fig.4}(b). 

    \par To investigate the effect of tensile strain on the critical temperature of germanene/Cr$_2$Ge$_2$Te$_6$ heterostructure, 
    biaxial tensile strain was applied. Strain is defined as $(a - a_0)/a_0$, where $a$ and $a_0$ are the lattice 
    constants with and without strain, respectively. As shown in Fig. \ref{Fig.4}(c), $T_C$ increases with strain. To address the 
    overestimation of $T_C$ by MC results, a rescaling method was employed. The calculated $T^{MC}_C$ of 50 K for ML 
    Cr$_2$Ge$_2$Te$_6$ without stain was scaled to match the experimental $T_C$ of 28 K for 2D Cr$_2$Ge$_2$Te$_6$ \cite{gong2017}. 
    All $T^{MC}_C$ values for strained germanene/ML Cr$_2$Ge$_2$Te$_6$ and ML Cr$_2$Ge$_2$Te$_6$ were then normalized by the 
    unstrained ML Cr$_2$Ge$_2$Te$_6$ $T^{MC}_C$ of 50 K. These ratios were subsequently multiplied by 28 K to provide more reliable 
    $T_C$ estimates as shown in Fig. \ref{Fig.4}(c). 

    \par
\begin{figure}[tphb]
    	\centering
        \includegraphics[scale=0.26]{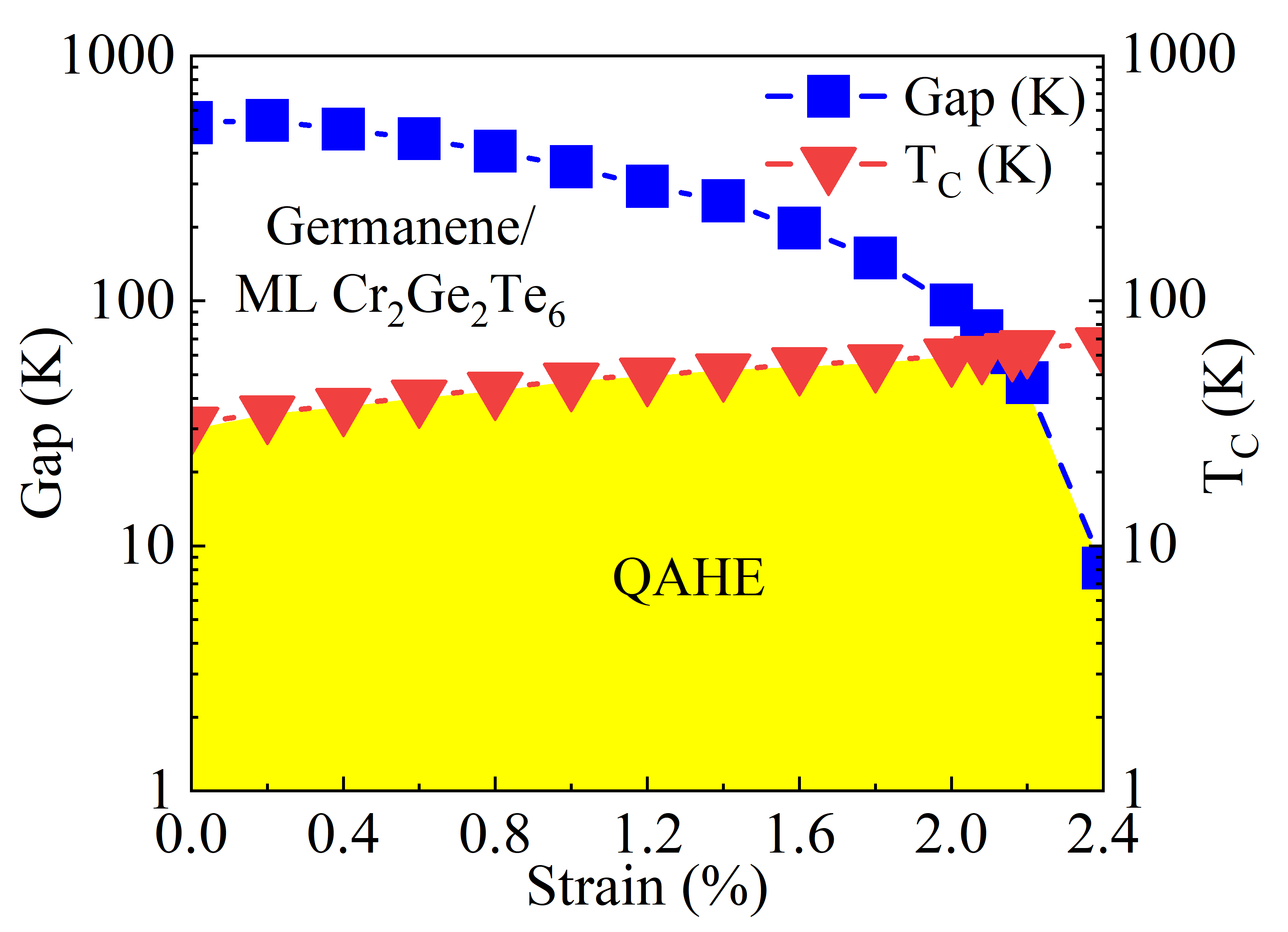}
        \caption{
    	The band gap and Curie temperature $T_C$ as a function of tensile strain of germanene/ML Cr$_2$Ge$_2$Te$_6$ 
        heterostructure. 
    	}\label{Fig.5}
    \end{figure}

 As shown in Fig. \ref{Fig.5}, increasing tensile strain in germanene/ML Cr$_2$Ge$_2$Te$_6$ heterostructure leads to 
    an increase in $T_C$ and a decrease in the band gap. The gap nearly closes at a tensile strain of 2.4\%. The band gap have 
    been given in unit of K. The yellow region in Fig. \ref{Fig.5} indicates the predicted QAHE phase. Notably, a QAHE state 
    is predicted up to 62 K in the germanene/ML Cr$_2$Ge$_2$Te$_6$ heterostructure with 2.1\% tensile strain.

    \par
\begin{figure}[tphb]
    	\centering
        \includegraphics[scale=0.3]{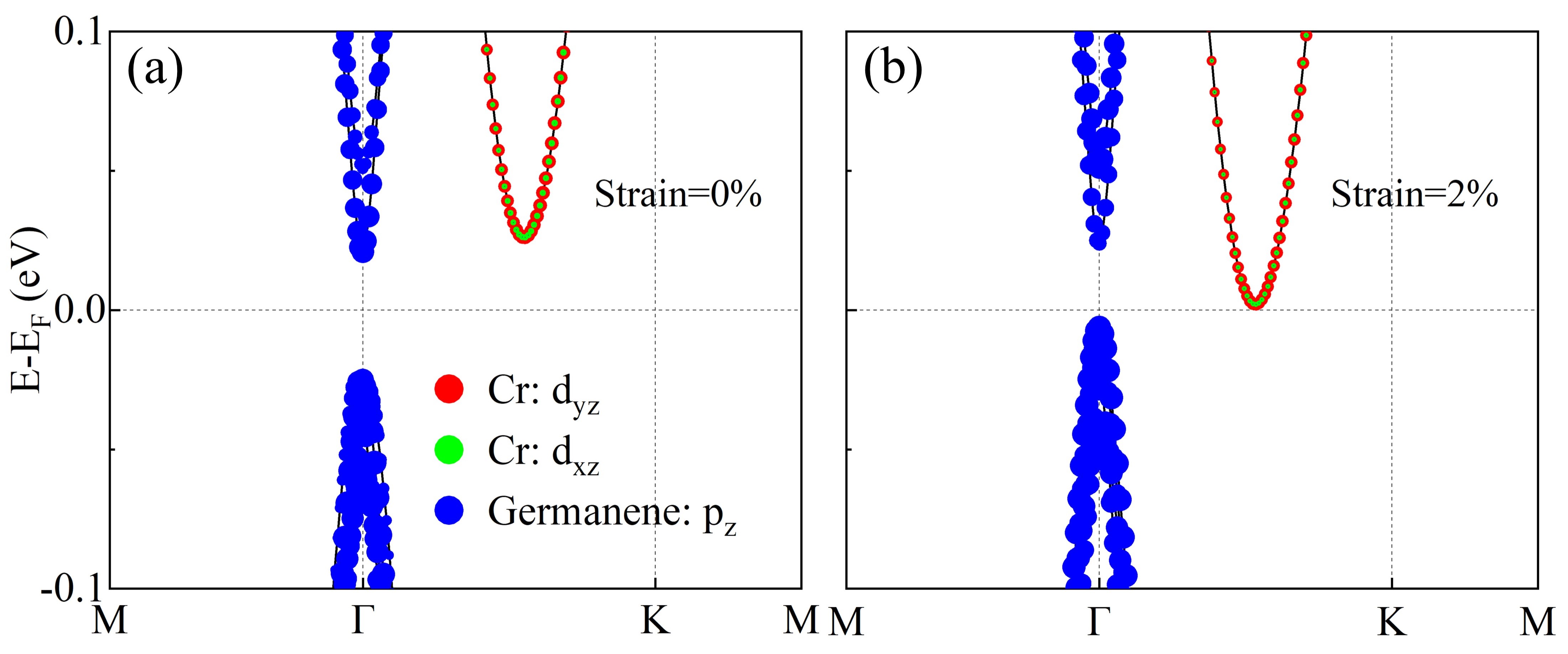}
        \caption{
    	The fatband around the Fermi level of germanene/ML Cr$_2$Ge$_2$Te$_6$ heterostructure (a) without strain and 
        (b) with 2\% tensile strain. 
    	}\label{Fig.6}
    \end{figure}

 To understand the decrease in band gap with increasing tensile strain, we analyzed the fatband structure around the 
    Fermi level for germanene/ML Cr$_2$Ge$_2$Te$_6$ heterostructure at strain=0\% and strain=2\% as shown in Fig. \ref{Fig.6}. 
    Clearly, the bands around $\Gamma$ point are mainly contributed by $p_z$ orbitals of germanene, while those along the $\Gamma - K$ 
    path are dominated by $d_{xz}$ and $d_{yz}$ orbitals of Cr. In the unstrained heterostructure, the direct band gap is located at $\Gamma$. 
    With increasing tensile strain, the bands at $\Gamma$ show little variation, but the bands along $\Gamma - K$ shift downwards, resulting 
    in a reduced band gap. 

    \par The great enhancement of $T_C$ in germanene/ML Cr$_2$Ge$_2$Te$_6$ heterostructure under tensile strain 
    can be explained by considering the superexchange interaction \cite{anderson1959,goodenough1955,kanamori1960}. 
    The superexchange interaction involves two processes. One is the direct exchange process between the $d_2$ electrons of the Cr$_2$ 
    atom and the $p'$ electrons of the Te atom, presented by $J_{p'd_2}$. The other is the hopping process between the $p$ electrons of 
    the Te atom and the $d_1$ electrons of the Cr1 atom, presented by $V_{p d_1}$. The coupling $J_{super}$ between the two Cr atoms 
    can be expressed as \cite{SMs} 
    
    \begin{equation}
        \begin{split}
            J_{super} &= \frac{1}{4} \sum_{d_1,p,p',d_2} |V_{p d_1}|^2 J_{p' d_2} 
            [\frac{1}{(E_{d_1 d_1'}^{\uparrow\uparrow})^2} - \frac{1}{(E_{d_1 d_1'}^{\uparrow\downarrow})^2}]. \\
            &= \frac{1}{4A} \sum_{d_1,p,p',d_2} |V_{p d_1}|^2 J_{p' d_2}, 
        \end{split}
        \label{equ3}
    \end{equation}

    \noindent where $A \equiv 1/(1/(E_{d_1 d_1'}^{\uparrow\uparrow})^2 - 1/(E_{d_1 d_1'}^{\uparrow\downarrow})^2)$ is taken as 
    a pending parameter and does not change with strain. $E_{d_1 d_1'}^{\uparrow\uparrow}$ and 
    $E_{d_1 d_1'}^{\uparrow\downarrow}$ are energies of two $d$ electrons at the same Cr atom with parallel and antiparallel spins, 
    respectively. $V_{p d_1}$ is the hopping parameter between the $p$ electrons and the $d_1$ electrons. The direct p-d exchange 
    $J_{p'd_2}$ can be derived from the s-d exchange model following the Schrieffer-Wolff transformation \cite{schrieffer1966} 
    
    \begin{equation}
        \begin{split}
            J_{p'd_2} = 2 |V_{p'd_2}|^2 (\frac{1}{E_{p'} - E_{d_2}} + \frac{1}{E_{d_2} + U - E_{p'}}), 
        \end{split}
        \label{equ4}
    \end{equation}

    \noindent where $E_{p'}$ and $E_{d_2}$ are the energy level of the $p'$ electrons and the $d_2$ electrons, respectively. 
    $V_{p'd_2}$ is the hopping parameter between the $p'$ electrons and the $d_2$ electrons. These parameters can be obtained 
    by DFT calculation and Wannier90 code. 

    \begin{table*}[!htbp]
        \centering
        \caption{$|V_{p d_1}|^2$ and $E_{p'} - E_{d_2}$ between $5p$ orbitals of Te 
        and $d_{z^2}$, $d_{x^2-y^2}$ orbitals of Cr for relaxed and strained ML Cr$_2$Ge$_2$Te$_6$ (in units of eV).}
        \vspace{5pt}
        \label{table1}
        \begin{tabular}{p{2cm} p{2cm} p{2cm} p{2cm} p{2cm} p{2cm} p{2cm}}\toprule
            &\multicolumn{6}{c}{Strain = 0\%} \\
            \cline{2-7}
            & $p_z - d_{z^2}$ & $p_x - d_{z^2}$ & $p_y - d_{z^2}$ & $p_z - d_{x^2-y^2}$ & $p_x - d_{x^2-y^2}$ & $p_y - d_{x^2-y^2}$ \\
            \hline
            $|V_{p d_1}|^2$ & 1.539 & 0.439 & 0.051 & 8.5E-5 & 0.026 & 0.005 \\
            %$J_{p' d_2}$ & 0.626 & 1.371 & 0.143 & 0.025 & 0.072 & 0.102 \\
            $E_{p'} - E_{d_2}$ & 0.430 & 0.551 & 0.415 & 0.317 & 0.438 & 0.302 \\
            \cline{2-7}
            $J_{super}A$ & \multicolumn{6}{c}{16.33} \\
            \midrule
            &\multicolumn{6}{c}{Strain = 2\%} \\
            \cline{2-7}
            & $p_z - d_{z^2}$ & $p_x - d_{z^2}$ & $p_y - d_{z^2}$ & $p_z - d_{x^2-y^2}$ & $p_x - d_{x^2-y^2}$ & $p_y - d_{x^2-y^2}$ \\
            \hline
            $|V_{p d_1}|^2$ & 1.407 & 0.425 & 0.065 & 1.5E-4 & 0.023 & 0.005 \\
            %$J_{p' d_2}$ & 0.973 & 4.517 & 0.082 & 0.026 & 0.051 & 0.734 \\
            $E_{p'} - E_{d_2}$ & 0.430 & 0.480 & 0.516 & 0.285 & 0.335 & 0.371 \\
            \cline{2-7}
            $J_{super}A$ & \multicolumn{6}{c}{18.43} \\
            \bottomrule
        \end{tabular}
    \end{table*}

    \par Due to the crystal field, the $d$ electron configuration of Cr is split into two groups: 
    $e_g$ ($d_{z^2}$ and $d_{x^2-y^2}$ orbitals) and $t_{2g}$ ($d_{xz}$, $d_{yz}$ and $d_{xy}$ orbitals). 
    Wannier90 calculations indicate that the main contribution comes from $e_g$. The values of $|V_{p d_1}|^2$ and 
    $E_{p'} - E_{d_2}$ for the Cr$_1$-Te$_1$-Cr$_2$ path in ML Cr$_2$Ge$_2$Te$_6$ without and with tensile strain are listed 
    in Table. \ref{table1}. %Every possible Te atoms with p orbitals are included to calculate $J_{12}$ by Equ. \ref{equ3}. 
    The resulting superexchange interaction $J_{super}$ for relaxed and strained ML Cr$_2$Ge$_2$Te$_6$ are $16.33/A$ and $18.43/A$, 
    respectively. The ratio of $J_{super}$ for strained to relaxed ML Cr$_2$Ge$_2$Te$_6$ is 1.13, which is comparable to the ratio 
    (1.25) of the DFT-calculated nearest neighbor exchange coupling $J_1$ for the strained and relaxed germanene/ML 
    Cr$_2$Ge$_2$Te$_6$ heterostructure. Comparison of the strained and unstrained ML Cr$_2$Ge$_2$Te$_6$ results reveals that the 
    enhancement of the superexchange interaction $J_{super}$ is primarily due to the decreased energy difference $E_{p'} - E_{d_2}$. 
    This decrease in $E_{p'} - E_{d_2}$ contributes to the enhancement of both $J_{super}$ and $T_C$. The Cr-Te-Cr bond angles are 
    89.8$^\circ$ and 91.2$^\circ$ for relaxed and strained ML Cr$_2$Ge$_2$Te$_6$, respectively, indicating no significant 
    difference in the hopping parameter $|V_{p d_1}|^2$ as shown in Table. \ref{table1}. 

    \par The application of strain could enhance the QAH transition temperature in the germanene/ML Cr$_2$Ge$_2$Te$_6$ 
    heterostructure. Because $T_C$ in 2D magnetic semiconductor Cr$_2$Ge$_2$Te$_6$ can be dramatically increased by strain, 
    which has been demonstrated in recent experiment \cite{oneill2023}. The fundamental mechanism lies in strain-induced reduction 
    of the energy difference between anions and cations, which contributes to the enhancement of $T_C$ by the superexchange mechanism. 
    However, the application of strain simultaneously reduces the band gap of the heterostructure germanene/ML Cr$_2$Ge$_2$Te$_6$ 
    by lowering the conduction bands. Consequently, the upper limit of the QAH temperature in germanene/ML Cr$_2$Ge$_2$Te$_6$ 
    heterostructure is determined by the intersection of $T_C$ and band gap curves. 

    \begin{figure}[tphb]
    	\centering
        \includegraphics[scale=0.48]{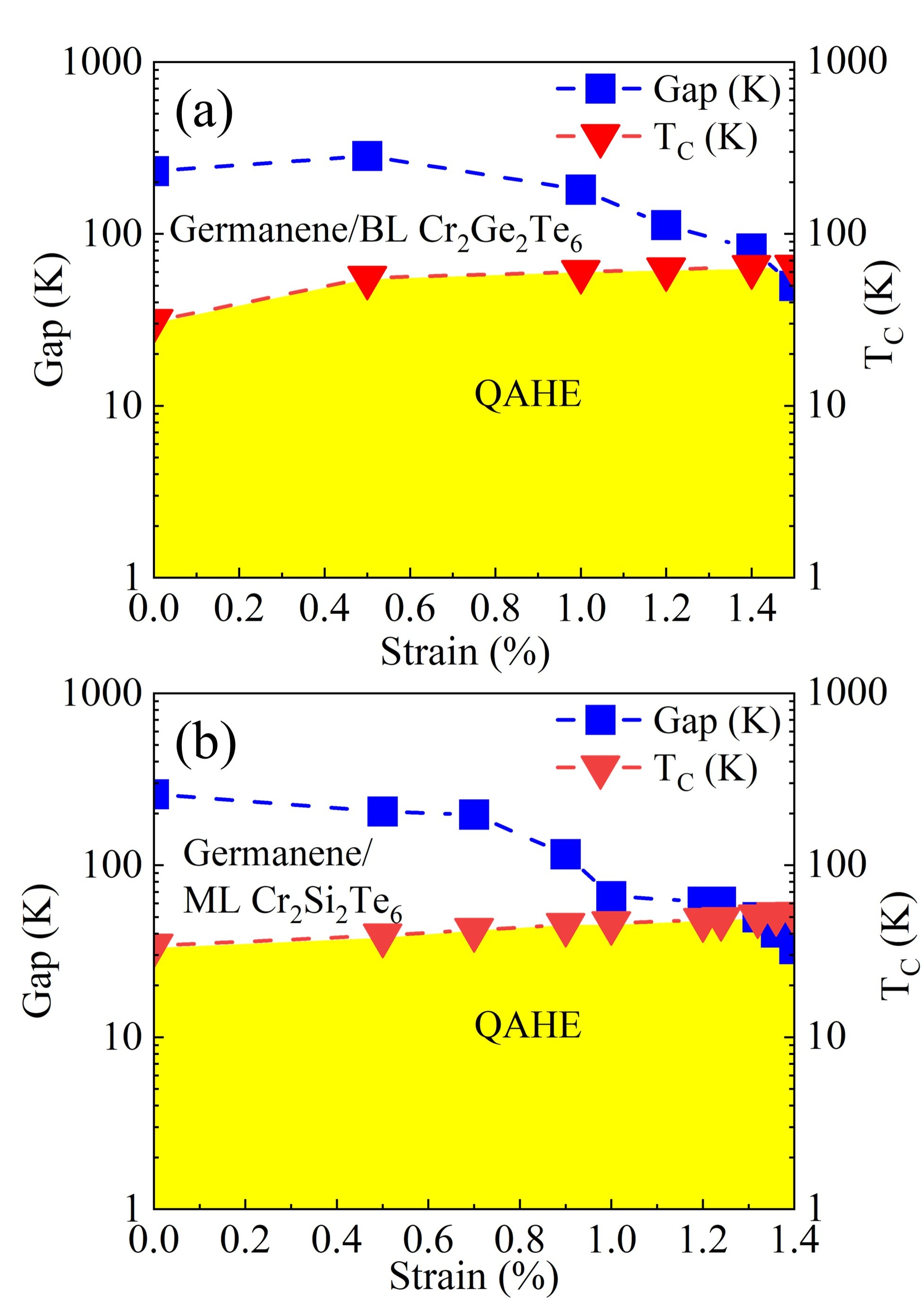}
        \vspace{-0.2cm}
        \caption{
    	The band gap and $T_C$ as a function of tensile strain of (a) germanene/BL Cr$_2$Ge$_2$Te$_6$ and 
        (b) germanene/ML Cr$_2$Si$_2$Te$_6$ heterostructures.
    	}\label{Fig.8}
    \end{figure}

    %\par The parameter A can be calculated in the following way. Two different spin configurations, FM and NAFM, are considered 
    %and the total energy can be expressed as 

    %\begin{equation}
    %    \begin{split}
    %        E_{FM} &= 12 J_1 |S|^2 + E_1 = 945/A + E_1 \\
    %        E_{NAFM} &= -12 J_1 |S|^2 + E_1 = -945/A + E_1 \\
    %    \end{split}
    %    \label{equ5}
    %\end{equation}

    %\noindent where $S$ is 3.62 $\mu_B$. For the FM spin configuration, the total energy is taken as $E_{FM} = 0$ as the 
    %energ reference. The total energy of NAFM configuration, $E_{NAFM} = 400$ meV, is derived form DFT results. 
    %By solving Eq. \ref{equ5}, the parameter A = -4731 meV$^2$ and $J_{12}$ = -1.270 meV. It is worth noting that the nearest 
    %neighbor exchange coupling $J_1$ given by DFT is -1.354 meV, which is very similar to $J_{12}$ given by superexchange model 
    %and Wannier90 code. 

    %%%%%%%%% Section4
    \section{QAHE in germanene/BL C\lowercase{R}$_2$G\lowercase{E}$_2$T\lowercase{E}$_6$ and germanene/ML C\lowercase{R}$_2$S\lowercase{I}$_2$T\lowercase{E}$_6$ heterostructures}

    \par We also investigated the QAHE in germanene/BL Cr$_2$Ge$_2$Te$_6$ and germanene/ML Cr$_2$Si$_2$Te$_6$ 
    heterostructures, and the results are shown in Fig. \ref{Fig.8}. Experiments have demonstrated an out-of-plane easy-axis and 
    interlayer ferromagnetic coupling in BL Cr$_2$Ge$_2$Te$_6$ \cite{gong2017}. To determine an appropriate Hubbard $U$ parameter, 
    we calculated the magnetic anisotropy energy (MAE) $(E_{\parallel}-E_{\perp})$ and the energy difference between AFM and FM 
    interlayer configurations $(E_{AFM}-E_{FM})$ as a function of $U$. As shown in SMs, a value of $U = 1.3$ eV accurately 
    reproduces the experimental observations for BL Cr$_2$Ge$_2$Te$_6$. Consequently, this value was adopted for germanene/BL 
    Cr$_2$Ge$_2$Te$_6$ calculations. The SMs also present the four different spin configurations of Cr atoms to calculate the 
    $T_C$ of BL Cr$_2$Ge$_2$Te$_6$. Our calculations reveal that the germanene/BL Cr$_2$Ge$_2$Te$_6$ heterostructure exhibits 
    the QAHE up to 64 K under a 1.4\% tensile strain as shown in Fig. \ref{Fig.8}(a). 

    \par For the germanene/ML Cr$_2$Si$_2$Te$_6$ heterostructure, a Hubbard $U$ value of 4 eV was employed in our calculations. 
    The similar rescale method is used to counteract the overestimation of $T_C$ by MC simulations. Specifically, the calculated 
    $T_C^{MC}$(Cr$_2$Si$_2$Te$_6$) of 27 K for ML Cr$_2$Si$_2$Te$_6$ was scaled to the experimentally determined $T_C^{exp}$ of 
    17 K for 2D Cr$_2$Si$_2$Te$_6$ \cite{zhang2021}. As shown in Fig. \ref{Fig.8}(b), the germanene/ML Cr$_2$Si$_2$Te$_6$ 
    heterostructure exhibits the QAHE up to 50 K under a 1.3\% tensile strain.

    %%%%%%%%% Section5
    \section{Discussion and Conclusion}
    \par The QAHE temperature of materials cannot exceed its $T_C$ and the topological band gap. For the 
    MnBi$_2$Te$_4$ with band gap of 100-200 meV ($\approx$ 1160-2320 K) and $T_C$ of 20 K, the upper limit of the 
    QAHE temperature cannot exceed 20 K. For the germanene/ML Cr$_2$Ge$_2$Te$_6$ heterostructure with biaxial 
    tensile strain, the upper limit of the QAHE temperature cannot exceed 62 K, as shown in Fig. \ref{Fig.5}. Thus, 
    the higher upper limit of QAHE temperature in germanene/ML Cr$_2$Ge$_2$Te$_6$ heterostructure is expected larger than 
    that in MnBi$_2$Te$_4$. 

    \par The realization of the room-temperature QAH state requires two fundamental criteria: (i) a 
    $T_C$ exceeding 300 K, and (ii) a topological band gap larger than 300 K ($\approx$ 26 meV). However, in practical 
    materials implementations, these ideal conditions often remain unattainable. A representative example is MnBi$_2$Te$_4$, 
    where the QAH effect was experimentally observed at cryogenic temperatures ($\sim$ 1-2 K), significantly below both 
    its intrinsic $T_C$ ($\sim$ 20 K) and the estimated band gap energy (100-200 meV) \cite{deng2020,liu2020,ge2020}. 
    The experimental observation deviates from theoretical predictions and can be attributed to two principal factors: 
    (1) extrinsic material imperfections such as structural disorders in samples, and (2) intrinsic magnetic fluctuations 
    occurring below $T_C$ that effectively suppress the QAH transition temperature. 

    \par The disorder effect can be categorized into two distinct types: magnetic disorder and nonmagnetic 
    disorder. Many studies have revealed that weak magnetic disorder stabilizes the edge states and the QAH conductance, 
    whereas strong disorder drives the system into an Anderson insulating phase \cite{chen2019,haim2019,lachman2015,lee2015,nomura2011a,qiao2016,wang2018}. 
    The nonmagnetic disorder induces layer-dependent topological phase transitions. Weak disorder generally enhances the 
    robustness of topological edge states and preserves the QAH conductance. By increasing disorder, the quantization will 
    also be disrupted \cite{okugawa2020,zhangChiral2021}. 

    \par We applied in-plane biaxial tensile strain that preserves the space group symmetry of the system. 
    The time-reversal symmetry is broken by the magnetic semiconductor in the heterostructure to realize the QAH state. 
    Strain can effectively enhance Curie temperature $T_C$ of some 2D ferromagnetic semiconductors, including Cr$_2$Ge$_2$Se$_6$ 
    \cite{dong2019}, MnSe$_2$ \cite{liP2024} and Fe$_2$Cl$_3$I$_3$ \cite{zhang2020}. When these materials form 
    heterostructures with topological insulators, it is highly expected that the QAH transition temperatures could be 
    synergistically enhanced due to the strain-enhanced $T_C$. 

    \par We note that the quantum spin Hall (QSH) effect at room temperature has been demonstrated in 
    Bi$_4$Br$_4$ \cite{shumiya2022}. In our paper, the QAH effect has been studied, which only has been observed in experiments 
    at very low temperature so far. According to the results in our paper, it will be very interesting to explore the possible 
    high temperature QAH effect in the heterostructures Bi$_4$Br$_4$/FS (FS = ferromagnetic semiconductors).  

    \par In this paper, we have investigated the strain-enhanced QAHE in germanene/ML Cr$_2$Ge$_2$Te$_6$, 
    germanene/BL Cr$_2$Ge$_2$Te$_6$ and germanene/ML Cr$_2$Si$_2$Te$_6$ heterostructures based on 
    first principle calculations. Our calculations demonstrate the emergence of topologically nontrivial edge states 
    and quantized anomalous Hall conductance in these heterostructures under tensile strain. We find that increasing 
    tensile strain leads to an increase in $T_C$ and a decrease in band gap within these germanene/magnetic 
    semiconductor heterostructures. Consequently, the highest QAHE temperatures achieved are 62 K for the germanene/ML 
    Cr$_2$Ge$_2$Te$_6$ heterostructure under 2.1\% tensile strain, 64 K for the germanene/BL Cr$_2$Ge$_2$Te$_6$ 
    heterostructure under 1.4\% tensile strain, and 50 K for the germanene/ML Cr$_2$Si$_2$Te$_6$ heterostructure under 
    1.3\% tensile strain. This work identifies experimental available materials for realizing high temperature QAHE states.

    \section*{Acknowledgements}
    This work is supported by National Key R\&D Program of China (Grant No. 2022YFA1405100), Chinese Academy of Sciences 
    (Grants No. YSBR-030, No. JZHKYPT-2021-08).

    %\newpage
    %\bibliographystyle{apsrev4-2}
    %\bibliography{ref1}
  

\end{document}